\documentclass[a4paper]{spicawStyle}
\usepackage{graphicx,natbib}

\begin{document}

\title{Monolithic Ge:Ga detector development for SAFARI}

\author{Y. Doi\inst{1} \and Y. Sawayama\inst{1} \and S. Matsuura\inst{2} \and M. Shirahata\inst{2} \and T. Arai\inst{2} \and S. Kamiya\inst{2} \and T. Nakagawa\inst{2} \and M. Kawada\inst{3}}

\institute{
The University of Tokyo, Komaba 3-8-1, Meguro, Tokyo, 153-8902, Japan
\and
Institute of Space and Astronautical Science, Japan Aerospace Exploration Agency, Yoshinodai 3-1-1, Sagamihara, Kanagawa, 229-8510, Japan
\and
Nagoya University, Furocho, Chikusa, Nagoya, 464-8601, Japan}

\maketitle 

\begin{abstract}

We describe the current status and the prospect for the development of monolithic Ge:Ga array detector for SAFARI.
Our goal is to develop a $64\times 64$ array for the 45 -- 110 $\mu$m band, on the basis of existing technologies to make $3\times 20$ monolithic arrays for the AKARI satellite.
For the AKARI detector we have achieved a responsivity of 10 A/W and a read-out noise limited NEP (noise equivalent power) of $10^{-17}$ W/$\sqrt{\rm Hz}$.
We plan to develop the detector for SAFARI with technical improvements; significantly reduced read-out noise with newly developed cold read-out electronics, mitigated spectral fringes as well as optical cross-talks with a multi-layer antireflection coat.
Since most of the elemental technologies to fabricate the detector are flight-proven, high technical readiness levels (TRLs) should be achieved for fabricating the detector with the above mentioned technical demonstrations.
We demonstrate some of these elemental technologies showing results of measurements for test coatings and prototype arrays.

\keywords{Instrumentation: detectors -- Infrared: general -- Missions: SPICA}
\end{abstract}

\section{Monolithic Ge:Ga Photoconductor Array}
\label{doiy2_sec:monolithic}
Ge:Ga photoconductor detector is a highly sensitive infrared detector that has been used for many astronomical applications including IRAS \citep{doiy2:IRAS}, IRTS (\citeauthor{doiy2:hiromoto92}, \citeyear{doiy2:hiromoto92}; \citeauthor{doiy2:shibai94}, \citeyear{doiy2:shibai94}), ISO (\citeauthor{doiy2:ISOPHOT}, \citeyear{doiy2:ISOPHOT}; \citeauthor{doiy2:ISOLWS}, \citeyear{doiy2:ISOLWS}), AKARI (\citeauthor{doiy2:doi02}, \citeyear{doiy2:doi02}; \citeauthor{doiy2:kawada07}, \citeyear{doiy2:kawada07}; \citeauthor{doiy2:shirahata09}, \citeyear{doiy2:shirahata09}), Spitzer \citep{doiy2:MIPS}, and Herschel \citep{doiy2:PACS}.

\begin{figure}[tb]
  \begin{center}
    \includegraphics[width=9 cm]{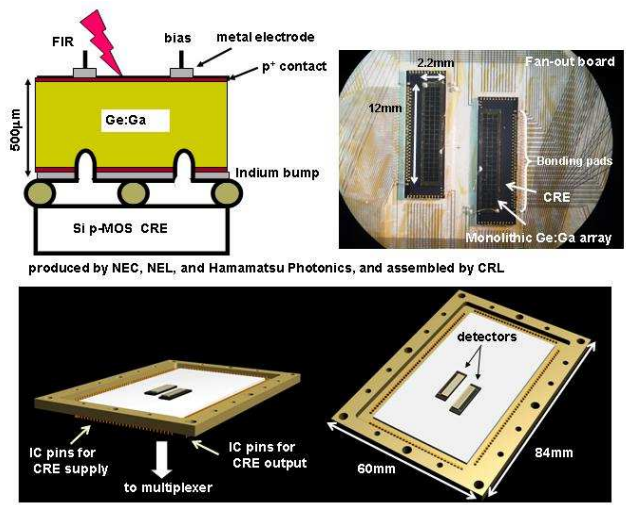}
  \end{center}
  \caption{Design of the $20\times 3$ pixel Ge:Ga monolithic array module onboard AKARI. Two sets of array are used for two photometric bands. Top-left figure shows a sectional view of the detector (see text for the detailed descriptions).}
\label{doiy2_fig:AKARI_SW}
\end{figure}

For the AKARI satellite, we have developed the world's first two-dimensional monolithic Ge:Ga photoconductor array, achieving a large array format of $20\times 3$ pixels with the pixel size of $500\times 500$ $\mu$m$^2$ (Figure~\ref{doiy2_fig:AKARI_SW}).
The Ge:Ga photoconductor is produced by doping a 0.5-mm-thick Ge wafer with Ga concentration of $1.6\times 10^{14}$ cm$^{-3}$.
For longitudinal configuration (top illumination), a transparent contact with a grid-shape metal electrode is fabricated on the top surface of the Ge:Ga wafer (also see $\S$\ref{doiy2_sec:transparent}).
Ohmic contact of the electrode is formed by B$^+$ ion implantation on top and bottom surfaces of the wafer.
Bias voltage is applied to the whole array in common from the top contact.
The bottom contact of each array pixel is directly connected to each CRE element with a 100-$\mu$m-diameter indium ball.
The array pixels are electrically separated each other by 50-$\mu$m-wide, 30-$\mu$m-deep grid-shape ditches on the back surface.

The infrared light incident from the top surface travels between the top and bottom surfaces back and forth, due to high reflectivity of the non-AR-coated top Ge surface and the metal back surface.
Thanks to the multiple reflections, long absorption length is obtained without any external optics like an optical cavity, and the resultant quantum efficiency is approximately 0.4, which corresponds to the responsivity at an optimum bias of $\sim 20$ A/W.
However, the multiple reflections cause interference fringes and result in the presence of periodic peaks in the spectral response, which is a serious drawback of this device especially for spectroscopic applications.
Mitigating this fringe problem by the anti-reflection (AR) coating is an issue of new array development (see $\S$\ref{doiy2_sec:AR}).

The arrays are attached to a fan-out board made of quartz substrate, and the fan-out board is mounted on a rigid frame made of gold-coated KOVAR.
These materials are chosen to minimize the thermal stress to the detector.
The KOVAR frame is mounted on the aluminium housing with thermal isolation legs made of Vespel SP-1 and is thermally connected to the helium tank by a copper strap.
The detector temperature is controlled to $\sim 2.0$ K by self-heat dissipation of the CRE ($\sim 1$ mW) and additional heater power.
The total mass of the detector module is $\sim 300$ g.
The flight model detector module passes the vibration test for simulating the mechanical stress at the launch. 

\section{New Array Design for SAFARI}
\label{doiy2_sec:new_array}
\begin{figure}[tb]
  \begin{center}
    \includegraphics[width=9 cm]{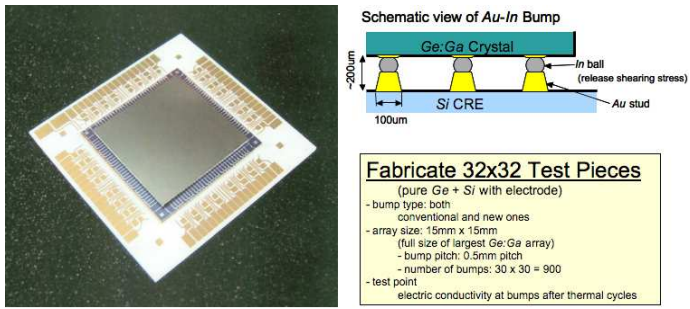}
  \end{center}
  \caption{A $32\times 32$ test array device to make proof of the Au-In bump technology for new design of Ge:Ga monolithic array for SAFARI.}
\label{doiy2_fig:test_array}
\end{figure}

The proposed design of monolithic array for SAFARI is a natural extension of the AKARI detector with minimum modification of fabrication technology.
We plan to make a $64\times 64$ mosaic array with four segments of $32\times 32$ array, which can be surely fabricated by using the same technology for the AKARI detector.
The pixel size of the SAFARI detector will be 500 $\times$ 500 $\mu$m$^2$, and the thickness will be 500 $\mu$m.
As shown in Figure~\ref{doiy2_fig:test_array}, we have already fabricated $32\times 32$ test arrays to make proof of the Au-In bump technology.
The test array consists of a pure Ge substrate and a Si board with $32\times 32$ bumping pads with pixel pitch of 500 $\mu$m, simulating the monolithic Ge:Ga array bumped on CRE.
We have checked the connection of the Au-In bumps of all pixels before and after thermal cycles between the room and LHe temperatures, by simple resistance measurements of the daisy chain electrodes.
For the initial test, we prepared a device with the indium balls smaller than that for the AKARI detector, to find out a minimum amount of indium for the bump.
As a result, the connection became failed at some pixels ($\sim 20$\%) after thermal cycles.
Hence, we are fabricating another test device with larger indium balls similar to the AKARI detector.
This device will be tested by the end of this year (2009).

\section{Transparent Contact}
\label{doiy2_sec:transparent}
A transparent contact is fabricated on the top surface of the Ge:Ga detector wafer by B$^+$ ion implantation (see $\S$\ref{doiy2_sec:monolithic}).
Investigation of the optimum implantation density is another development item for the SAFARI detector as it may cause additional absorption and/or reflection.

\begin{figure}[tb]
  \begin{center}
    \includegraphics[width=5 cm]{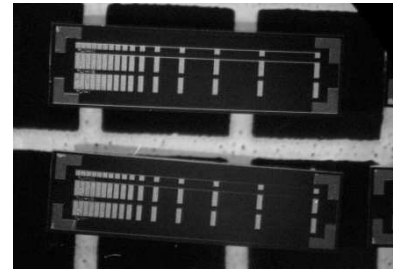}
  \end{center}
  \caption{Test pieces for the resistivity measurement by transmission line method (TLM).}
\label{doiy2_fig:tlm_photo}
\end{figure}

\begin{figure}[tb]
  \begin{center}
    \includegraphics[width=7 cm]{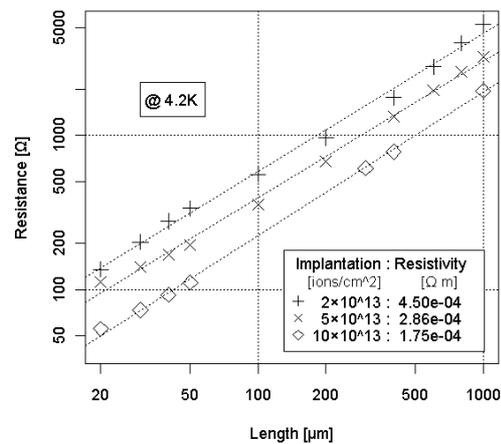}
  \end{center}
  \caption{Resistivity measurement of the transparent contact.}
\label{doiy2_fig:tlm}
\end{figure}

For that purpose, we have tested three parameters including $2,\ 5,\ 10\times 10^{13}\ {\rm [ions\ cm^{-2}]}$, which are 1/2, 1, and 2 times of the implantation for the AKARI detector \citep{doiy2:fjw00}.
A figure of test pieces for the resistivity measurement is shown in Figure~\ref{doiy2_fig:tlm_photo} and the results are shown in Figure~\ref{doiy2_fig:tlm}.
We found reasonably small resistivity ($<5\times 10^4\ {\rm [\Omega\ m]}$) for all the parameters with a linear dependence for the implantation density (Figure~\ref{doiy2_fig:tlm}).

\begin{figure}[tb]
  \begin{center}
    \includegraphics[width=7 cm]{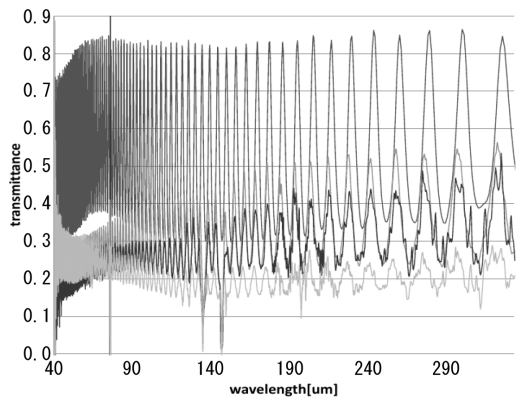}
  \end{center}
  \caption{Transmittance of transparent contacts as well as a pure Ge wafer for a reference. The top data are the transmittance of pure Ge and the bottom three data are that of contacts on Ge:Ga wafer. The implantation density is $2,\ 5,\ 10\times 10^{13}\ {\rm [ions\ cm^{-2}]}$ from top to bottom. Implantation energy is $40\ {\rm [keV]}$. All the substrates are 500 $\mu$m in thickness. Measurements are done under 4.2 [K].}
\label{doiy2_fig:contact_trans}
\end{figure}

We also measure the optical transmittance of the contacts and the results are shown in Figure~\ref{doiy2_fig:contact_trans}.
Transmittance of 0.2 -- 0.4 are measured for all the transparent contacts on Ge:Ga wafers with some dependence upon the implantation density.
These transmittance are significantly smaller than that of the pure Ge wafer shown in Figure~\ref{doiy2_fig:contact_trans} and
the degradations cannot be fully attributed to Ga and B free careers in Ge:Ga substrate and the contacts at the LHe temperature.
Part of the degradation could possibly be due to metallization of the contact.
We plan to make further investigations to find out the optimum parameters for the contact fabrication.

\section{Anti-Reflection Coating}
\label{doiy2_sec:AR}

In order to avoid the interference fringe problem described in $\S$\ref{doiy2_sec:monolithic}, top surface of the monolithic array for SAFARI is AR-coated by Si/SiO2 multi-layer coatings.
Even if the optical path length inside Ge:Ga is decreased by the AR-coatings, the mean optical efficiency of the coated device, taking the reflection loss at the initial surface into account, can be higher than that without coatings.

\begin{figure}[tb]
  \begin{center}
    \includegraphics[width=7 cm]{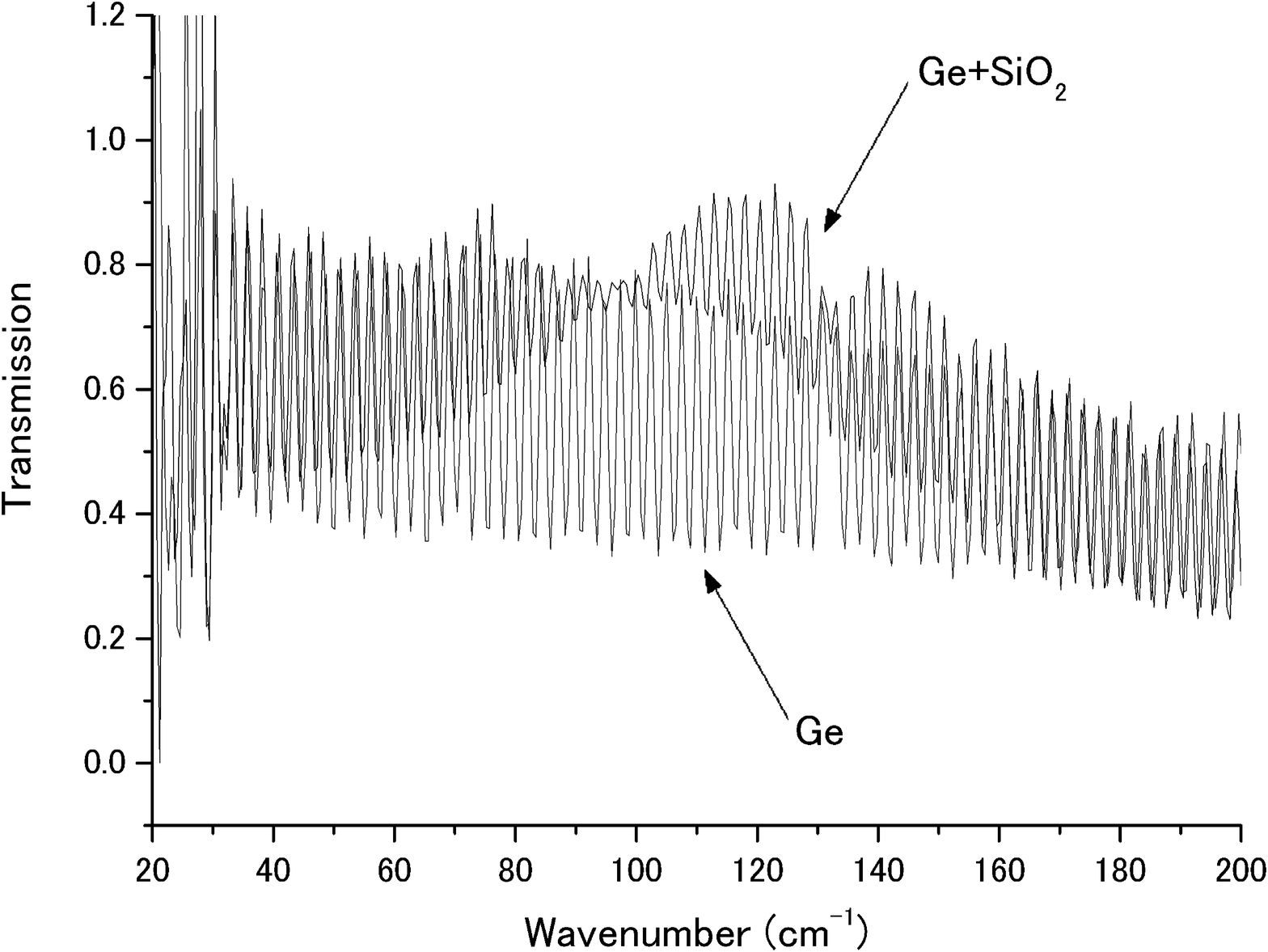}
    \includegraphics[width=6.5 cm]{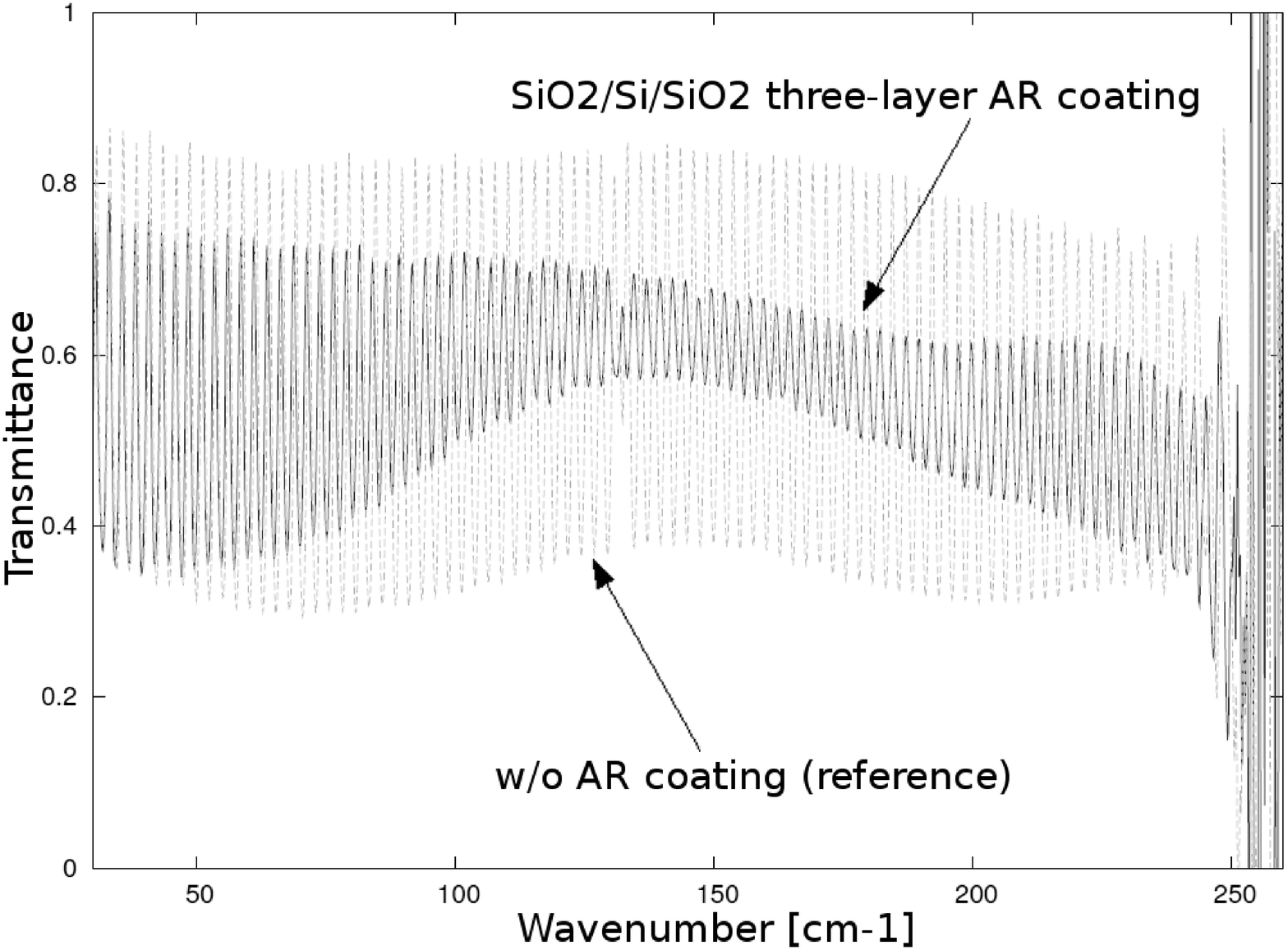}
  \end{center}
  \caption{The transmittance measurements for the AR-coated Ge samples. Top panel: a single layer ${\rm SiO_2}$ coating. Bottom panel: three layers (${\rm SiO_2/Si/SiO_2}$) coatings.}
\label{doiy2_fig:ar}
\end{figure}

We have already made samples of the AR-coatings for transmission measurements and the results are shown in Figure~\ref{doiy2_fig:ar}.
The initial trial of the process was a single layer ${\rm SiO_2}$ on a Ge wafer, and we could successfully demonstrate that the interference fringes around the center wavelength disappear and that the transmittance showed good agreement with the theoretical calculation.
As the next step, a sample of three layer coatings (${\rm SiO_2/Si/SiO_2}$) was made to demonstrate the broad-band AR coatings, and its broad band property was measured.
Although the minimum reflectance of the three-layer coatings was much larger than that of the single layer and the transmittance showed considerable amount of residual interference fringes around the center wavelength, the effective bandwidth of the reduced reflectance was much wider than that of the single layer.
Comparison of the data with theoretical calculation is currently in progress and we plan to make further development of the AR-coating with design refinement as well as increased number of layers.
We have also fabricated the AR-coatings on single element Ge:Ga chips and on the $5\times 5$ prototype arrays, and we will demonstrate the improved properties of these devices in near future.

\section{Prototype $5\times 5$ Array}
\label{doiy2_sec:doiy2_5x5}

\begin{figure}[tb]
  \begin{center}
    \includegraphics[width=9 cm]{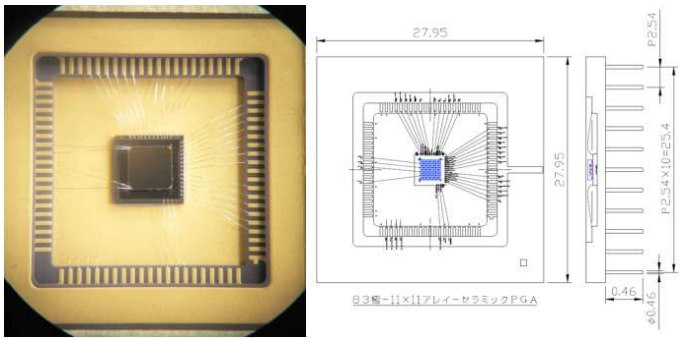}
  \end{center}
  \caption{A $5\times 5$ prototype array mounted on a chip carrier with PGA to demonstrate the properties of the monolithic array.}
\label{doiy2_fig:5x5}
\end{figure}

In order to demonstrate the properties of the monolithic array, we have fabricated a prototype Ge:Ga array of $5\times 5$ pixels bumped to a cryogenic CTIA readout circuit with a $5\times 5$ multiplexer.
For the first trial fabrication, commercially purchased Ge:Ga material is used.
The device mounted on a chip carrier with pin grid array (PGA) is shown in Figure~\ref{doiy2_fig:5x5}.

Structure of the flight model array will be basically the same as this device except for the array size.
The IC pins of the chip carrier are connected to a socket of a fan-out board for interfacing to the back stage.

\begin{figure}[tb]
  \begin{center}
    \includegraphics[width=6 cm]{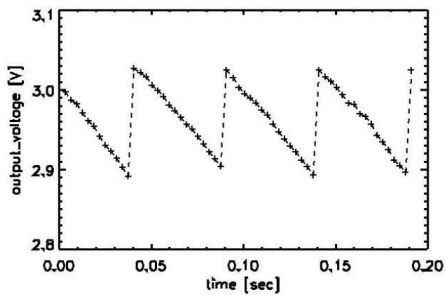}
  \end{center}
  \caption{Output signal of a detector pixel under illumination of far-infrared light. Periodical reset of integrated careers is applied at every 50 [msec].}
\label{doiy2_fig:ramp}
\end{figure}

\begin{figure}[tb]
  \begin{center}
    \includegraphics[width=6.4 cm]{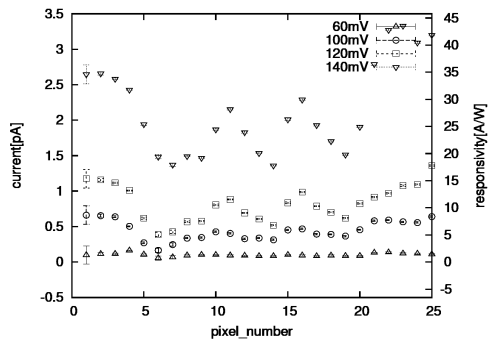}
  \end{center}
  \caption{Photo-current of the $5\times 5$ pixels and its dependence to the bias voltage.}
\label{doiy2_fig:photo_current}
\end{figure}

A preliminary operation test has been carried out and the results are shown in Figures~\ref{doiy2_fig:ramp} and \ref{doiy2_fig:photo_current}.
Figure~\ref{doiy2_fig:ramp} demonstrates an example signal output of a pixel illuminated by low-temperature black-body light installed in a test cryostat.
Photo-current integrated in a capacitor together with periodical reset of the integrated electrical charge results in a saw-like ramp signal.
Currently environmental EMC noise dominates the output noise performance so that a modification of the test environment is required.
Figure~\ref{doiy2_fig:photo_current} shows the measured responsivity of each pixel in the $5\times 5$ array and its dependence on the applied bias voltage.
An average responsivity of $\sim 5$ [A/W] with 30\% variation is found with a bias voltage of 100 [mV].
This responsivity is somewhat lower than that has been achieved for the AKARI array ($\S$\ref{doiy2_sec:monolithic}).
At the same time, a periodical variation pattern across the array is found.

We will make further investigations on the detector performances with optimising contact process parameters as well as changing detector material to the Ge:Ga crystal that was used for AKARI flight module.

\section{Scope of the Development}
\label{doiy2_sec:schedule}

\begin{figure}[tb]
  \begin{center}
    \includegraphics[width=9 cm]{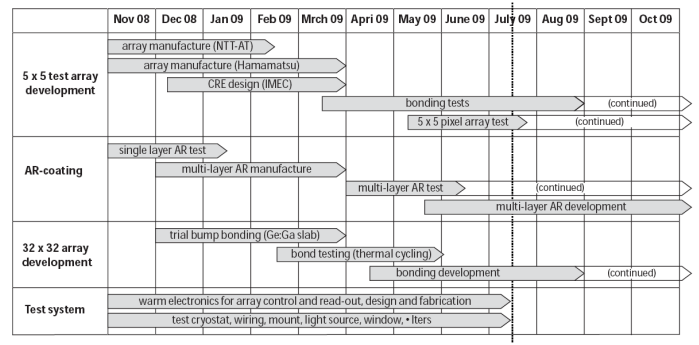}
  \end{center}
  \caption{Monolithic Array Development Timeline.}
\label{doiy2_fig:timeline}
\end{figure}

We are developing a $64 \times 64$ Ge:Ga array as a candidate detector array for SAFARI.
For that purpose, we are aiming to demonstrate the feasibility of all the elemental technologies that are required for fabricating a Ge:Ga array by mid 2010 (Figure~\ref{doiy2_fig:timeline}).
Now we start testing a prototype $5 \times 5$ array and plan to demonstrate fundamental detector performances including responsivity and noise performance with an improved test environment.
Another trial fabrication of $5 \times 5$ arrays is ongoing and an array with AKARI material as well as AR coated detector are to be fabricated.
With these devices, we plan to measure the best performance that can be achieved with the current cold readout electronics.
Mitigation of fringe problem ($\S$\ref{doiy2_sec:monolithic}), including fringes in spectral response as well as an optical cross-talk, is another issue to be tested with the new device with AR coating.
Indium bumping with larger indium balls ($\S$\ref{doiy2_sec:new_array}) is planned to achieve $32 \times 32$ direct bumping.
AR coating will also be developed with design refinement and increased number of layers ($\S$\ref{doiy2_sec:AR}).
All these developments are ongoing and are expected to fulfil the requisite detector performance in an appropriate time frame.

\end{document}